\documentclass[12pt]{article}
\usepackage{scicite}
\usepackage{graphicx}
\usepackage{amssymb}
\usepackage{times}
\newenvironment{sciabstract}{%
\begin{quote} \bf}
{\end{quote}}

\newcounter{lastnote}

\def\gr{$\gamma$-ray}
\def\ES{1ES~1101-232}
\def\ESS{1ES~0229+200}
\def\H{H~2356-309}
\def\ESSS{1ES~0347-121}

\title{Evidence for strong extragalactic magnetic fields from Fermi observations of TeV  blazars} 

\author
{Andrii Neronov$^{1\ast}$ and Ievgen Vovk,$^{1}$\\
\\
\normalsize{$^{1}$Data Centre for Astrophysics (ISDC), Geneva Observatory,}\\
\normalsize{Ch. d'Ecogia 16, Versoix,1290, Switzerland}\\
\\
\normalsize{$^\ast$E-mail:  Andrii.Neronov@unige.ch.}
}

\date{}

\begin{document} 
\baselineskip24pt
\maketitle 

\begin{sciabstract}
Magnetic fields in galaxies are produced via the amplification of seed magnetic fields of unknown nature. The seed fields, which might  exist in their initial form in the intergalactic medium, were never detected. We report a lower bound $B\ge 3\times 10^{-16}$~gauss on the strength of intergalactic magnetic fields, which stems from the nonobservation of GeV gamma-ray emission from electromagnetic cascade initiated by tera-electron volt gamma-ray in intergalactic medium. The bound improves as $\lambda_B^{-1/2}$  if magnetic field correlation length, $\lambda_B$, is much smaller than a megaparsec. This lower bound constrains models for the origin of cosmic magnetic fields.
\end{sciabstract}

The problem of the origin of 1- to 10-$\mu$G magnetic fields in galaxies and galaxy clusters is one of the long-standing problems of astrophysics and cosmology [see \cite{kronberg94,grasso01,widrow02,kulsrud08} for reviews]. It is assumed that the observed magnetic fields result from the amplification of much weaker seed fields. However, the nature of the initial weak seed fields is largely unknown. There are two broad classes of models for the seed fields: astrophysical models, which assume that the seed fields are generated by motions of the plasma in (proto)galaxies, and cosmological models in which the seed fields are produced in the early universe \cite{kronberg94,grasso01,widrow02,kulsrud08}. 

Extremely weak unamplified extragalactic magnetic fields (EGMFs) have escaped detection up to now.  Measurements of the Faraday rotation in the polarized radio emission from distant quasars \cite{kronberg94,blasi,vallee04} and/or distortions of the spectrum and polarization properties in the cosmic microwave background (CMB) radiation \cite{barrow97,subramanian98,durrer00,jedamzik00,seshadri01,mack02,lewis04,yamazaki08,giovannini08,kristiansen08,kahniashvili08,seshadri09} imply upper limits on EGMF strengths at the level of $\sim 10^{-9}$~G.  Numerical modeling of magnetic field formation in galaxy clusters implies a theoretical upper bound of the order of $\sim 10^{-12}$~G on EGMF strength \cite{dolag99,dolag05}.  Bounds on the EGMF strength depend on the field correlation length $\lambda_B$, which is also unknown. A lower limit on $\lambda_B$ is set by the requirement that the resistive magnetic diffusion time scale has to be larger than the age of the Universe \cite{grasso01}, whereas an upper limit is set only by the size of the visible part of the Universe, $R_H$.

Here we report a lower bound for the EGMF strength, derived from the data of Fermi and High Energy Stereoscopic System (HESS) gamma-ray telescopes. Similarly to the existing upper bounds, the lower bound depends on the unknown EGMF configuration parameters, such as the typical correlation length and spectrum.

Gamma rays with energies above $\sim 1$~TeV cannot propagate over cosmological distances because of absorption resulting from interactions with diffuse extragalactic background light (EBL) \cite{gould67b,1ES_nature,0229_HESS,franceschini08}. The mean free path of gamma rays of energy $E_{\gamma_0}$ through EBL is $D_\gamma\simeq 80\kappa\left(E_{\gamma_0}/10\mbox{ TeV}\right)^{-1}$~Mpc, where $\kappa\sim 1$ is a numerical factor that accounts for uncertainties of the measurements and modeling of the EBL \cite{neronov09}. Interactions of multi-TeV gamma rays with the EBL lead to the deposition of electron-positron pairs in the intergalactic space. These $e^+e^-$ pairs emit secondary cascade gamma rays via Inverse Compton (IC) scattering of CMB photons. Typical energies for the IC photons emitted by electrons of energy $E_e\simeq E_{\gamma_0}/2$ are $E_{\gamma} = (4/3)\epsilon_{CMB}(E_e/m_ec^2)^2\simeq 88 \left[E_{\gamma_0}/10 \mbox{ TeV}\right]^2\mbox{ GeV}$, where $\epsilon_{CMB}=6\times 10^{-4}$~eV is the typical energy of CMB photons and $E_e$ and $m_e$ are the energy and mass, respectively, of an electron. Pairs lose energy on IC scattering  on the distance scale $D_e\simeq 10^{23}(E_e/10\mbox{ TeV})^{-1}$~cm, which is much smaller than the gamma ray mean free path $D_\gamma$. Power removed from the primary gamma-ray beam is transferred to the cascade gamma-ray emission.

If magnetic fields, which deviate electron and positron trajectories, are negligibly small, the IC emission from the electromagnetic cascade contributes to the primary point gamma-ray source flux \cite{plaga,fan02,davezac07,murase08,aharonian}. Otherwise, if magnetic fields along the path of development of the cascade are strong enough to deviate the trajectories of the pairs, the cascade emission appears as extended emission around the initial point source \cite{coppi,neronov07,neronov09,elyiv09,dolag09}.

The deflection angle $\delta$ depends on the correlation length of the magnetic field.  If $\lambda_B\gg D_e$, the motion of electrons or positrons can be approximated by the motion in a homogeneous magnetic field. In this case $\delta\simeq D_e/R_L \simeq 3\times 10^{-4}\left[B/10^{-16}\mbox{ G}\right]\left[E_e/10\mbox{ TeV}\right]^{-2}$ is a ratio of $D_e$ to the  Larmor radius $R_L$. If $\lambda_B\ll D_e$, electron deflections are describable by diffusion in angle, so that the deflection angle is $\delta=\sqrt{D_e\lambda_B}/R_L \simeq 5\times 10^{-5} \left[E_e/10\mbox{ TeV}\right]^{-3/2} \left[B/10^{-16}\mbox{ G}\right] \left[\lambda_B/1\mbox{ kpc}\right]^{1/2}$. The size of the extended cascade source is estimated as $\Theta_{\rm ext}\simeq \delta/\tau$, where $\tau=D/D_\gamma$ is the optical depth for gamma rays from a source at a distance $D$ with respect to absorption on EBL \cite{neronov09}.

Because lower energy electrons are deviated by larger angles, the size of the extended cascade source $\Theta_{\rm ext}$ is larger at low energies. The energy of cascade photons, $E_{\gamma,min}$, below which the extended source size becomes larger than the point spread function (PSF) of a telescope, depends on the EGMF strength and correlation length. 
In the case of the Fermi telescope, the PSF depends on the photon energy, 
decreasing as $\Theta_{\rm PSF}\simeq 2^\circ\left[E_\gamma/1\mbox{ GeV}\right]^{-0.8}$ ($95\%$ of the signal) below $E_\gamma\simeq 1$~GeV and improving from $\sim 2^\circ$ at $1$~GeV to $\Theta_{\rm PSF}\simeq 0.2^\circ$ at $E_\gamma\sim 10$~GeV\cite{fermi_psf}. Taking the photon energy $E_\gamma\simeq 10$~GeV as a reference, one finds that $\Theta_{\rm ext}\ge \Theta_{\rm PSF}$ if
\begin{equation}
\label{Bpsf}
B\ge B_{PSF}\simeq\left\{
\begin{array}{ll}
6\times 10^{-17}\tau\left[E_{\gamma,min}/ 10\mbox{ GeV}\right]\mbox{ G}, & \lambda_B>D_e\\
8\times 10^{-16}\tau\left[E_{\gamma,min}/ 10\mbox{ GeV}\right]^{3/4}\left[\lambda_{B}/1\mbox{ kpc}\right]^{-1/2}\mbox{ G}, & \lambda_B<D_e
\end{array}
\right.
\end{equation}

To constrain the presence of a cascade contribution in the spectra of distant TeV blazars, we have analyzed Fermi/LAT (Large Array Telescope) data on  the blazars \ES, \ESS, \ESSS\ and \H, obtained during the Fermi's first year of operation. These sources were selected on the basis of their high redshifts ($z=0.14$ for \ESS, $z=0.165$ for \H, $z=0.186$ for \ES\ and $z=0.188$ for \ESSS) and hard TeV band spectra \cite{1ES_nature,0229_HESS,H2356_HESS,1ES_HESS,1ESSS_HESS,note_other_results}. 

\begin{figure}
\includegraphics[width=12cm]{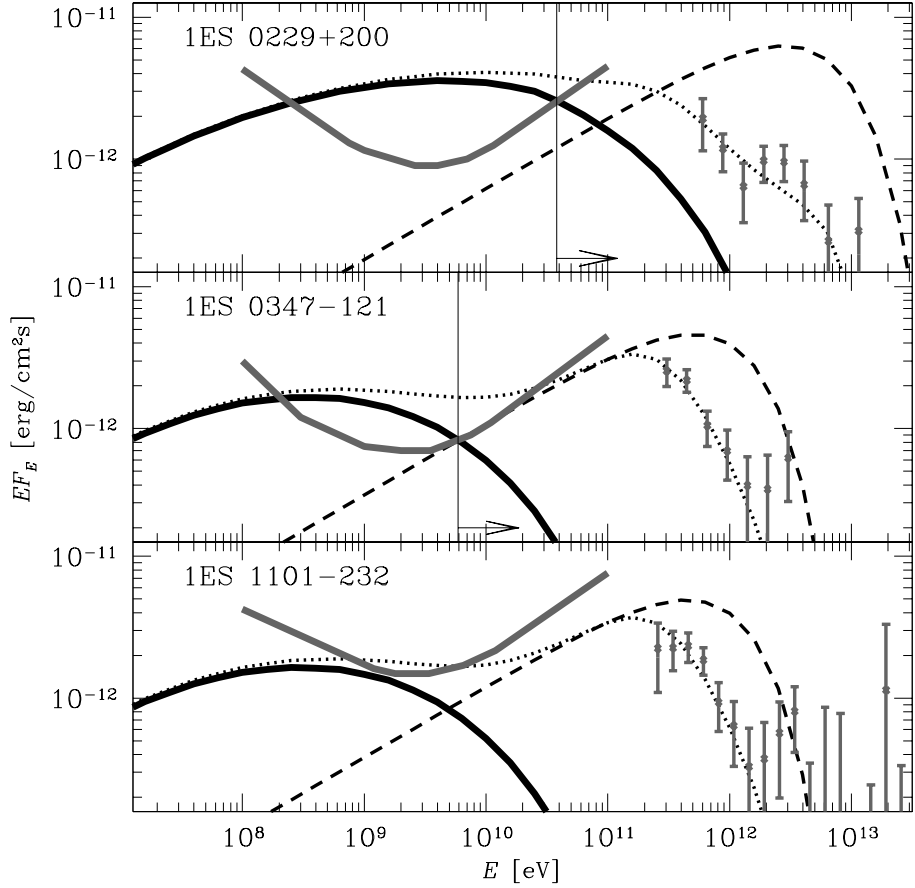}
\caption{A comparison of models of cascade emission from TeV blazars (thick solid black curves) with Fermi upper limits (grey curves) and HESS data (grey data points). Thin dashed curves show the primary (unabsorbed) source spectra. Dotted curves show the spectra of electromagnetic cascade initiated by pair production on EBL. Vertical lines with arrows show the energies below which the cascade emission should be suppressed.}
\label{fig:spectra}
\end{figure}

Minimal possible cascade signal from \ES\ and \H\ are below the Fermi upper bounds [see supporting online material (SOM) for details].
The nondetection of \ESS\ and \ESSS\ with Fermi imposes restrictions on the contribution to the flux from the  low-energy tail of the cascade \cite{note_h2356}.
In order to determine these restrictions, we have developed a numerical model of electromagnetic cascade in intergalactic space, initiated by the absorption of TeV gamma rays on EBL photons. The model solves the one-dimensional kinetic equations for concentrations of gamma rays and of $e^+e^-$ pairs, taking into account the injection of pairs by the absorbed gamma rays \cite{gould67}, the injection of secondary cascade gamma rays via IC emission by the $e^+e^-$ pairs and the cooling of the $e^+e^-$ pairs via IC scattering energy losses \cite{blumenthal70}. The cosmological photon backgrounds with which our gamma rays and $e^+e^-$ pairs interact include CMB and EBL taken from \cite{franceschini08}. 

 The initial gamma-ray spectra at the source, shown by dashed thin curves in Fig. \ref{fig:spectra}, were chosen in the form of a cutoff powerlaw, $dN_\gamma/dE\sim E^{-\Gamma}\exp(-E/E_{cut})$. The cascade emission power is equal to the fraction of the power of the primary gamma-ray beam absorbed on the way from the source to Earth. Because almost 100\% of the power initially injected at the energies above TeV is absorbed, the luminosity of cascade emission is roughly equal to the integral primary source luminosity in the multi-TeV energy band.  Primary source spectra shown in Fig. \ref{fig:spectra} correspond to the minimal value of $E_{cut}$ compatible with HESS data, to minimize total flux of the cascade contribution (see details in the SOM). 

Assumption of zero magnetic field along the cascade development path is in contradiction with Fermi upper bounds on the source fluxes (Fig. \ref{fig:spectra}).  Although the model spectra deviate from simple powerlaws, the deviations are small, meaning that Fermi bounds on the powerlaw-type spectra could be applied.

The cascade emission has to be suppressed below an energy $E_{\gamma,min}$ (marked by a vertical line in the three graphs of Fig. \ref{fig:spectra}) at which the Fermi upper bound becomes higher than the model cascade flux. Suppression of the cascade flux at low energies could be achieved if trajectories of low-energy $e^+e^-$ pairs are deviated by magnetic fields. The cascade emission does not contribute to the point source flux below the energy $E_{\gamma,min}$ if EGMF is stronger than $B_{\rm PSF}$ given by Eq. 1. The values of $B_{\rm PSF}$ corresponding to $E_{\gamma,min}$ found for each source are given in table S1.

\begin{figure}
\includegraphics[width=12cm]{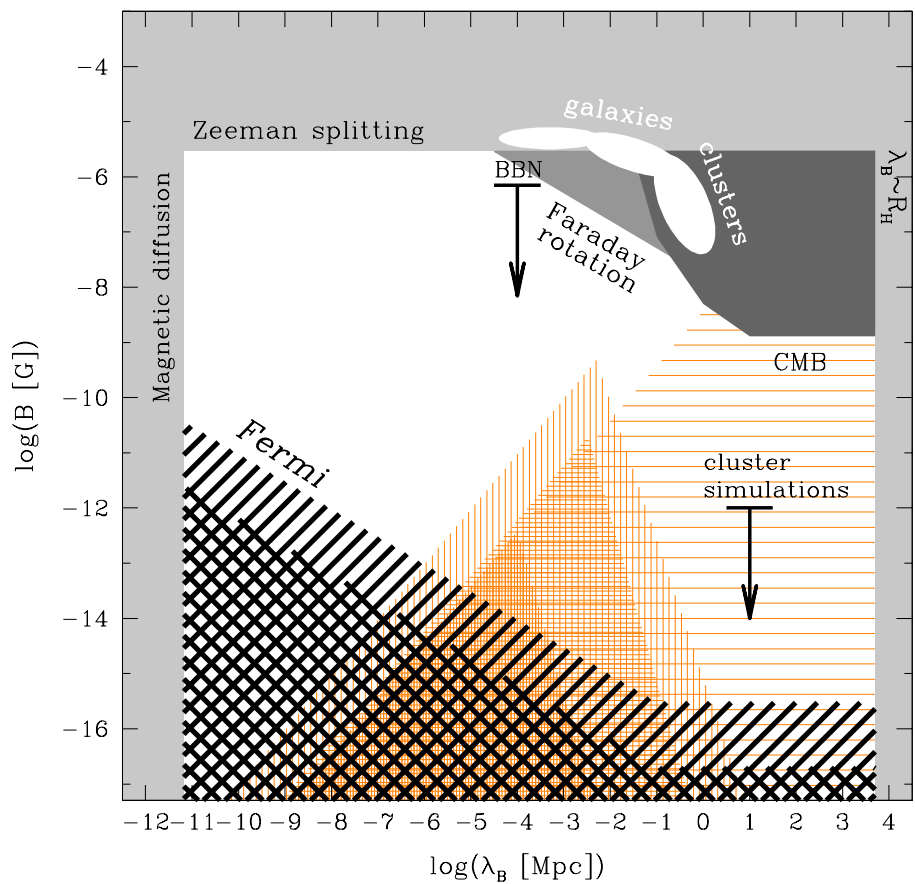}
\caption{Light, medium and dark grey: known observational bounds on the strength and correlation length of EGMF, summarized in the Ref. \cite{neronov09}. The bound from Big Bang Nucleosynthesis marked ``BBN" is from the Ref. \cite{grasso01}. The black hatched region shows the lower bound on the EGMF derived in this paper. Orange hatched regions show the allowed ranges of  $B, \lambda_B$ for magnetic fields generated at the epoch of Inflation (horizontal hatching) the electroweak phase transition (dense vertical hatching), QCD phase transition (medium vertical hatching), epoch of recombination (rear vertical hatching) \cite{neronov09}.  White ellipses show the range of measured magnetic field strengths and correlation lengths in galaxies and galaxy clusters.}
\label{fig:exclusion}
\end{figure}

The best bound $B\ge B_{\rm PSF}$, imposed by Fermi limits on the flux from \ESS\ (black hatched region in Fig. \ref{fig:exclusion}), suffers from a number of uncertainties and, therefore, should be considered as an order-of-magnitude estimate. 

Suppression of the cascade contribution to the point source flux below the energy $E_{\gamma,min}$ results in a deviation of the model source flux from the powerlaw at the energies $E\lesssim E_{\gamma,min}$. This means that the Fermi upper bounds on the powerlaw-type spectra shown in Fig. \ref{fig:spectra} could provide only rough estimates of $E_{\gamma,min}$ and $B_{\rm PSF}$.  Additional uncertainty is introduced in the estimate of $B_{\rm PSF}$ by the uncertainty of the measurements of EBL which result in the uncertainty of the optical depth $\tau$ in Eq. 1. Further uncertainty is introduced in the derivation of $E_{\gamma,min}$ from non-simultaneous data in GeV and TeV bands. Both HESS and Fermi measurements refer to the source spectra averaged over year(s) time scale. The reported HESS observations of the sources took place in the period from 2005 to 2006 \cite{1ES_nature,0229_HESS,1ES_HESS}, whereas Fermi measurements were taken in 2008 and 2009.  Up to now, no long-term variability was found in HESS observations \cite{1ES_nature,0229_HESS,1ES_HESS,1ES_2008}.

The mean free path of primary multi-TeV gamma rays is of the order of $\sim 80\kappa$~Mpc. The largest structures in the universe, galaxy clusters, have typical sizes of the order of several Mpc and their volume filling factor is small. Most of the volume of the sphere of radius $\sim 80\kappa$~Mpc around \ES, \ESS\ and \ESSS\ is occupied by the voids in the Large-Scale Structure.

Evidence for existence of magnetic fields in the voids provides a strong argument in favor of a cosmological origin of the fields serving as seeds for subsequent amplification in galaxies and galaxy clusters. Weak magnetic fields produced in the early universe are expected to fill the whole universe, including the voids. Contrarily, in the astrophysical models, the weak seed fields are created locally in (proto)galaxies and the field outside these structures should be close to zero. 

Cosmological magnetogenesis models consider generation of magnetic fields with a correlation length that does not exceed the size of the cosmological horizon and with energy density that does not exceed the critical density of the universe at the moment of magnetogenesis.  Four broad classes of cosmological magnetogenesis scenaria are considered: magnetogenesis at the epoch of inflation, at the electroweak phase transition, at the epoch of quantum chromodynamics (QCD) phase transition and at the epoch of recombination \cite{kronberg94,grasso01,widrow02,hogan83,turner88,quashnock89,vachaspati91,ratra92,sigl97,diaz08,demozzi09}. The lower bound reported here excludes substantial parts of allowed parameter space for all the classes of cosmological magnetogenesis models (Fig. \ref{fig:exclusion}).

\section*{Supporting online material}

The {\it Fermi}/LAT data were filtered using the {\it gtselect} tool. Spectral analysis was done with the help of the likelihood technique using the {\it gtlike} tool, as explained in the {\it Fermi}/LAT data analysis threads \cite{analysis}. We have included point sources visible in {\it Fermi}/LAT images, Galactic diffuse emission and isotropic diffuse \gr\ emission components to the \gr\ emission model for the likelihood analysis. The {\it mapcube} file \textit{gll\_iem\_v02.fits} was used for the Galactic diffuse emission modeling, together with a corresponding tabulated model for the isotropic diffuse emission. The likelihood analysis for each object was performed inside a circular region with an angular radius of 14 degrees, centered on  the source. 

 To derive an upper limit on the source fluxes from the LAT data, we applied the following procedure.  Assuming that the source flux in the 0.1-100~GeV band is a powerlaw
$dN_\gamma/dE=N_0\left(E/0.1\mbox{ GeV}\right)^{-\Gamma}$ we found the upper bound  on normalization factor, $N_{0}$, for each $\Gamma$. To do this, we  calculated the Test-Statistics (TS) \cite{mattox} value as a function of $N_0$ and $\Gamma$. For each fixed value of $\Gamma$ we plotted the TS value as a function of flux normalization to derive the 95\% confidence level upper bound $N_{0,95}(\Gamma)$ on the normalization factor, following procedure of Ref. \cite{mattox}. The resulting upper limits on the source fluxes shown as grey curves in Fig. 1 are the envelopes of the entire set of spectra  
$\left. dN_\gamma\right/dE_\gamma=N_{0,95}(\Gamma)\left(\left. E_\gamma\right/ 0.1\mbox{ GeV}\right)^{-\Gamma}$. 

\begin{figure}
\includegraphics[width=\linewidth]{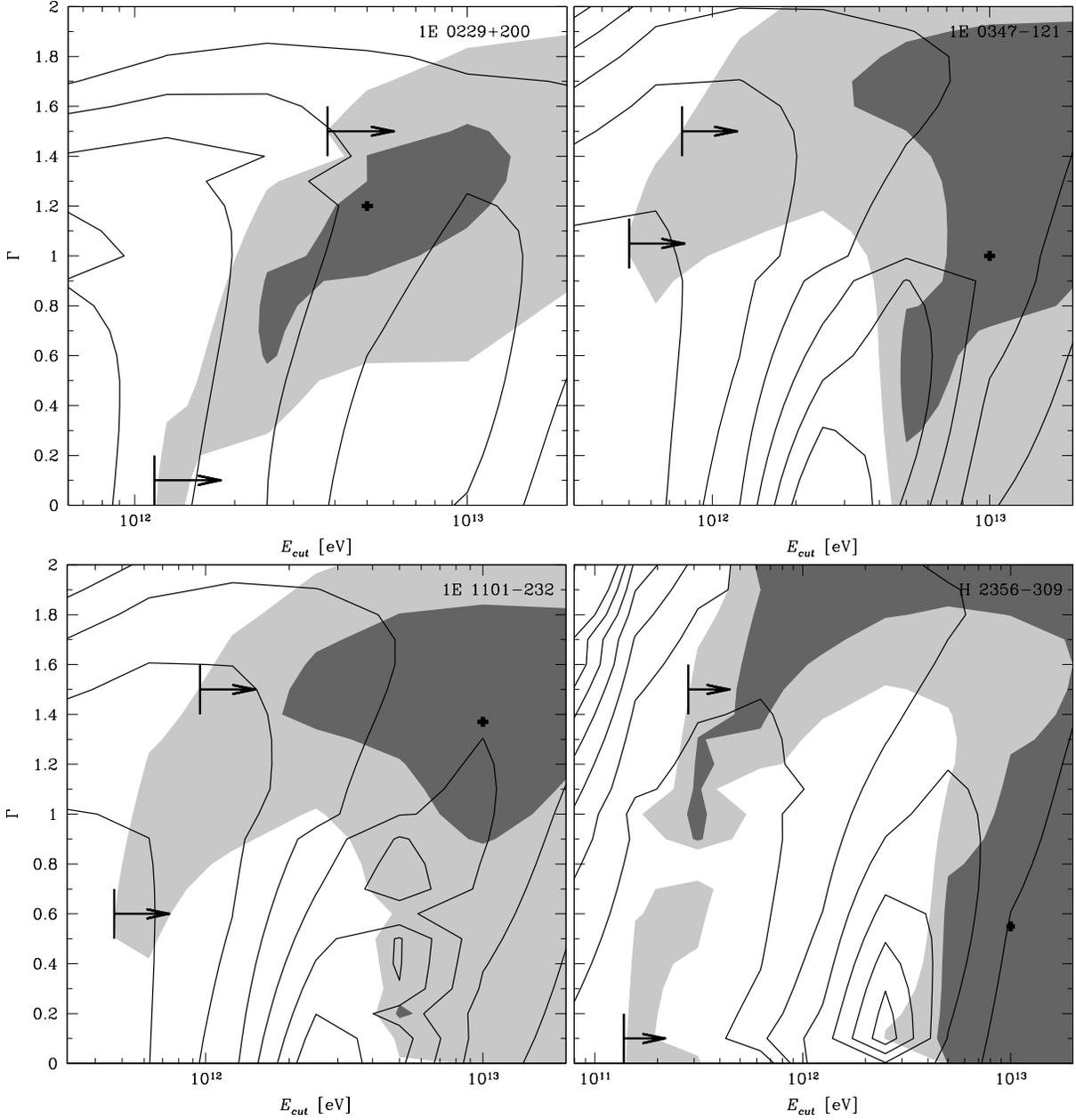}
\caption{Greyscale: 68 and 95\% confidence levels for the cut-off energy $E_{cut}$ and photon index $\Gamma$ found from the fitting of HESS spectra. Crosses mark the best-fit parameter values. Contours show the levels of integral energy flux above 0.1~TeV for the fitted spectra at each $E_{cut}$ and $\Gamma$, with the increments of $2.5\times 10^{-12}$~erg/cm$^2$s. Arrows show the 95\% confidence level lower bounds on the cut-off energy for the photon index $\Gamma\ge 1.5$ (upper arrows) and absolute lower bounds on $E_{cut}$ (lower arrows). }
\label{fig:fits}
\end{figure}

In order to calculate the bounds on the parameters of initial source spectra, we have fitted HESS data points in the 0.1-10~TeV energy band with model spectra for different $E_{cut}$ and $\Gamma$, calculated  under the assumption of zero magnetic field along the line of sight. For each model spectrum with given $E_{cut}, \Gamma$ we found the normalization by fitting the model to the HESS data.  Next, for each pair $E_{cut}, \Gamma$ we found the $\chi^2$ of the fit and the total \gr\ flux emitted at the energies $E\ge 0.1$~TeV. The dependence of the $\chi^2$ and of the total flux on $E_{cut},\Gamma$ for the model fits of the HESS data is shown in Fig. \ref{fig:fits}.

To find the values of parameters of initial source spectrum which minimize the cascade contribution to the source flux at zero EGMF strength, we chose a pair of cut-off energy $E_{cut,min}$ and powerlaw index $\Gamma$ lying in the 95\% confidence contour and corresponding to the minimal integral energy flux above $0.1$~TeV. 
We consider the values of $\Gamma$ softer than $\Gamma=1.5$, a restriction commonly adopted in the modeling of blazar spectra in the very high-energy \gr\ band, see e.g. Refs. {\it (22-24)}. We have verified that the results of our analysis (the existence of a lower bound on the EGMF strength) do not change if we relax the constraint $\Gamma\ge 1.5$ (see Table S1). The lower bounds on the cut-off energy found from the analysis of HESS spectra are given in Table S1, together with the estimates of $E_{\gamma,min}$ and $B_{\rm PSF}$.

\begin{table}
\begin{tabular}{|r|rrr|rrr|}
\hline
&&$\Gamma\ge 1.5$ &&&all $\Gamma$&\\
\hline
Source & $E_{cut,min}$ &$E_{\gamma,min}$ &$B_{\rm PSF} (\lambda\gg D_e)$& $E_{cut,min}$ &$E_{\gamma,min}$&$B_{\rm PSF} (\lambda\gg D_e)$\\
\hline
1ES 0229+200 & 3.8 TeV & 39 GeV & $3\times 10^{-16}$~G &1.2 TeV & 30 GeV&$3\times 10^{-16}$~G \\
1ES 0347-121 & 0.8 TeV & 6 GeV & $2\times 10^{-17}$~G &0.5 TeV & 6 GeV&$2\times 10^{-17}$~G \\
1ES 1101-232 & 1.0 TeV & - &  - &0.5 TeV & - & - \\
H 2356-309 & 0.3 TeV & - & - &0.14 TeV & -&- \\
\hline
\end{tabular}
\label{tab:params}
\caption{Parameters of the model spectra and limits on EGMF for the analyzed sources.}
\end{table}


\begin{thebibliography}{99}
\bibitem{kronberg94}  P.P.~Kronberg, {\it Rept.\ Prog.\ Phys.}  {\bf 57}, 325 (1994).
\bibitem{grasso01} D.~Grasso D., H.R.~Rubinstein, {\it Phys. Rept.}  {\bf 348}, 163 (2001).
\bibitem{widrow02} L.~M.~Widrow,  {\it Rev. Mod. Phys.}  {\bf 74}, 775 (2002).
\bibitem{kulsrud08} R.~M.~Kulsrud, E.~G.~Zweibel, {\it Rept. Prog. Phys.}  {\bf 71}, 0046901(2008).
\bibitem{blasi}  P.~Blasi, S.~Burles and A.~V.~Olinto,  {\it Astrophys. J.}  {\bf 514}, L79 (1999). 
\bibitem{vallee04} J.P.Vallee, {\it N. Astron. Rev.} {\bf 48}, 763 (2004). 
\bibitem{barrow97} J.D.~Barrow, P.G.~Ferreira, J.~Silk, {\it Phys. Rev. Lett.} {\bf 78}, 3610 (1997).
\bibitem{subramanian98} K.~Subramanian, J.D.~Barrow, {\it Phys. Rev. Lett.} {\bf 81}, 3575 (1998).
\bibitem{durrer00} R.~Durrer, P.G.~Ferreira, T.~Kahniashvili, {\it Phys. Rev.  D} {\bf 61}, 043001 (2000).
\bibitem{jedamzik00}  K.~Jedamzik, V.~Katalinic and A.V.~Olinto,  {\it Phys. Rev. Lett.}  {\bf 85}, 700 (2000).
\bibitem{seshadri01} T.R.~Seshadri, K.~Subramanian, {\it Phys. Rev. Lett.} {\bf 87}, 101301 (2001).
\bibitem{mack02} A.~Mack,  T.~Kahniashvili, A.~Kosowsky, {\it Phys. Rev. D} {\bf 65}, 123004 (2002).
\bibitem{lewis04}  A.~Lewis, {\it Phys. Rev. D} {\bf 70}, 043011 (2004).
\bibitem{yamazaki08} D.G.~Yamazaki {\it et al.}, {\it Phys. Rev. D} {\bf 77}, 043005 (2008).  
\bibitem{giovannini08} M.~Giovannini, K.E.~Kunze, {\it Phys. Rev. D} {\bf 77}, 063003 (2008). 
\bibitem{kristiansen08} J.R.~Kristiansen, P.G.Ferreira, {\it Phys. Rev. D} {\bf 77}, 123004 (2008).   
\bibitem{kahniashvili08} T.~Kahniashvili, T.~Maravin, A.~Kosowsky,  {\it Phys. Rev.  D} {\bf 80}, 023009 (2008). 
\bibitem{seshadri09} T.R.~Seshadri, K.~Subramanian, {\it Phys. Rev. Lett.} {\bf 103}, 081303 (2009).
\bibitem{dolag99} K.~Dolag, M.~Bartelmann, H.~Lesch, {\it Astron. Astrophys.} {\bf 348}, 351 (1999).
\bibitem{dolag05} K.~Dolag, D.~Grasso, V.~Springel, I.~Tkachev, {\it J. Cosmol. Astroparticle Phys.} {\bf 0501}, 009 (2005).
\bibitem{gould67b} R.J.~Gould, G.P.~Schr\'eder, {\it Phys. Rev. Lett.} {\bf 16}, 252 (1967).
\bibitem{1ES_nature} F.A.~Aharonian, {\it et al.}, {\it Nature} {\bf 440},  1018 (2006).
\bibitem{0229_HESS} F.A.~Aharonian, {\it et al.}, {\it Astron. Astrophys.} {\bf 475}, L9 (2007).
\bibitem{franceschini08} A.~Franceschini, G.~Rodighiero, M.~Vaccari, {\it Astron. Astrophys.} {\bf 487}, 837 (2008).
\bibitem{neronov09} A.~Neronov, D.~Semikoz, {\it Phys. Rev. D.} {\bf 80}, 123012 (2009).
\bibitem{plaga}  R.~Plaga, {\it Nature} {\bf 374}, 430 (1995).
\bibitem{fan02} Y.Z.~Fan, Z.G.~Dai, D.M.~Wei, {\it Astron. Astrophys.} {\bf 415}, 483 (2002).
\bibitem{davezac07}  P.~D'Avezac, G.~Dubus, B.~Giebels, {\it Astron. Astrophys.} {\bf 469}, 857 (2007). 
\bibitem{murase08} K.~Murase, {\it et al.}, {\it Astrophys. J.}, {\bf  686} L67 (2008). 
\bibitem{coppi} F.~A.~Aharonian, P.~S.~Coppi, H.~J.~Volk,  {\it Astrophys. J.} {\bf 423}, L5 (1994). 
\bibitem{neronov07} A.~Neronov, D.V.~Semikoz, {\it J. Exp. Theor. Phys. Lett.} {\bf 85}, 473, (2007).
\bibitem{elyiv09} A.~Elyiv, A.~Neronov, D.V.~Semikoz, {\it Phys. Rev. D} {\bf 80}, 023010 (2009). 
\bibitem{dolag09} K.~Dolag, M.~Kachelriess, S.~Ostapchenko, R.~Tomas, {\it Astrophys. J.} {\bf 703}, 1078 (2009).
\bibitem{fermi_psf} {\tt http://www-glast.slac.stanford.edu/software/IS/\\ glast\_lat\_performance.htm}
\bibitem{H2356_HESS} F.A.~Aharonian, {\it et al.}, {\it Astron. Astrophys.} {\bf 455}, 461 (2006).
\bibitem{1ES_HESS} F.A.~Aharonian, {\it et al.}, {\it Astron. Astrophys.} {\bf 470}, 475 (2007).
\bibitem{1ESSS_HESS} F.A.~Aharonian {\it et al.}, {\it Astron. Astrophys.} {\bf 473}, L25 (2007).

\bibitem{note_other_results} When this paper was ready for publication, we became aware of analysis of the signal from TeV blazars  reported by the Fermi collaboration \cite{abdo09}. Our analysis is consistent with the one of \cite{abdo09} and extends the results obtained in \cite{abdo09}. 

\bibitem{abdo09} A.A.~Abdo {\it et al.},  {\it Astrophys. J.} {\bf 707}, 1310 (2009).

\bibitem{note_h2356} The cut-off energies of intrinsic spectra of \H\ and \ES\ could be so low that the flux of the cascade emission could be below Fermi upper bounds on GeV source flux. This is illustrated in the bottom graph of Fig. \ref{fig:spectra} for the case of \ES. See SOM for more details.

\bibitem{gould67} R.J.~Gould, G.P.~Schr\'eder, {\it Phys. Rev.} {\bf 155}, 1404 (1967).
\bibitem{blumenthal70} G.R.~Blumenthal, R.J.~Gould,  {\it Rev. Mod. Phys.} {\bf 42}, 237 (1970).
\bibitem{1ES_2008} F.A.~Aharonian, {\it et al.}, {\it Astron. Astrophys.} {\bf 478}, 387 (2008).
\bibitem{turner88} M.S.~Turner, L.M.~Widrow, {\it Phys. Rev. D} {\bf 37}, 2743 (1988).
\bibitem{quashnock89} Quashnock, J., A. Loeb, D. Spergel,  {\it Astrophys. J. Lett.}  {\bf 344}, L49 (1989).
\bibitem{hogan83} C.J.~Hogan, {\it Phys. Rev. Lett.} {\bf 51}, 1488 (1983).
\bibitem{vachaspati91} T. Vachaspati, {\it Phys. Lett. B} {\bf  265}, 258 (1991).
\bibitem{ratra92} B.~Ratra {\it Astrophys. J.} {\bf 391}, L1 (1992).
\bibitem{sigl97} G.~Sigl, A.V.~Olinto, K.~Jedamzik, {\it Phys. Rev. D} {\bf 55}, 4582 (1997).
\bibitem{diaz08} A.~Diaz-Gil, J.~Garcia-Bellido, M.G.~Perez, A.~Gonzalez-Arroyo, {\it Phys. Rev. Lett.} {\bf 100}, 241301 (2008).
\bibitem{demozzi09} V.~Demozzi, V.~Mukhanov, H.~Rubinstein, {\it J. Cosmol. Astroparticle Phys.} {\bf 08}, 025 (2009).  

\bibitem{analysis} {\tt http://fermi.gsfc.nasa.gov/ssc/data/analysis/scitools/}.
\bibitem{mattox} J.R.~Mattox et al., {\it Ap.J.}, {\bf 461}, 396 (1996).
\bibitem{aharonian} F.~A.~Aharonian, A.~N.~Timokhin and A.~V.~Plyasheshnikov, {\it A\&A}, {\bf 384}, 834 (2002).
\end{thebibliography}
\end{document}